\documentclass[useAMS,usenatbib,usegraphicx]{mn2e}
\usepackage{graphicx, txfonts}

\defcitealias{paper1}{Paper~1}
\defcitealias{paper2}{Paper~2}

\title[Transit Variability in Bow Shock-Hosting Planets]{Transit Variability in Bow Shock-Hosting Planets}
\author[Vidotto, Jardine, Helling]{A.~A.~Vidotto\thanks{E-mail: Aline.Vidotto@st-andrews.ac.uk},  M.~Jardine, Ch.~Helling\\
SUPA, School of Physics \& Astronomy, University of St Andrews, North Haugh, St Andrews,  KY16 9SS, UK\\
}
\begin{document}

\date{Accepted 2011 February 6.  Received 2011 February 1; in original form 2010 December 17}

\pagerange{\pageref{firstpage}--\pageref{lastpage}} \pubyear{2011}

\maketitle

\label{firstpage}

\begin{abstract}
We investigate the formation of bow shocks around exoplanets as a result of the interaction of the planet with the coronal material of the host star, focusing on physical causes that can lead to temporal variations in the shock characteristics. 
We recently suggested that WASP-12b may host a bow shock around its magnetosphere, similarly to the one observed around the Earth. For WASP12b, the shock is detected in the near-UV transit light curve. Observational follow-up suggests that the near-UV light curve presents temporal variations, which may indicate that the stand-off distance between the shock and the planet is varying. This implies that the size of the planet's magnetosphere is adjusting itself in response to variations in the surrounding ambient medium. 
We investigate possible causes of shock variations for the known eccentric ($e>0.3$) transiting planets. We show that, because the distance from the star changes along the orbit of an eccentric planet, the shock characteristics are modulated by orbital phase. During phases where the planet lies inside (outside) the corotation radius of its host star, shock is formed ahead of (behind) the planetary motion. We predict time offsets between the beginnings of the near-UV and optical light curves that are, in general, less than the transit duration. Variations in shock characteristics caused in eccentric systems can only be probed if the shock is observed at different orbital phases, which is, in general, not the case for transit observations. However, non-thermal radio emission produced by the interaction of the star and planet should be modulated by orbital phase.  
We also quantify the response of the shock to variations in the coronal material itself due to, e.g., a non-axisymmetric stellar corona, planetary obliquity (which may allow the planet to move through different regions of the host star's corona), intrinsic variations of the stellar magnetic field (resulting in stellar wind changes, coronal mass ejections, magnetic cycles). Such variations do not depend on the system eccentricity. We conclude that, for systems where a shock is detectable through transit light curve observations, shock variations should be a common occurrence.
\end{abstract}

\section{Introduction}
More than 100 exoplanets have now been detected through optical transit observations. Transit light curves can provide a wealth of information about both the object in transit and its host star, such as the planet radius $R_p$, orbital period $P_{\rm orb}$, orbital semi-major axis $a$, stellar mass $M_*$ and radius $R_*$ \citep{seager2003}. Observations of light curves in wavelengths other than in the optical should also provide more information about the system. This is the case of the gas giant planet WASP-12b, whose near-UV light curve differs from the optical one by presenting an earlier beginning \citep{fossati2010}. This early ingress has been interpreted as due to the presence of material around the planet, although several explanations for the origin of this material have been suggested in the literature (\citealt{lai2010}; \citealt*{paper1}, hereafter Paper~1). In \citetalias{paper1}, we suggested that this material probes the existence of a bow shock around the planet, which is formed when the relative motion between the planet and the stellar coronal material is supersonic. A bow shock is observed around the Earth's magnetosphere as a result of the interaction of our planet with the solar coronal wind. According to our model, the detection of the shock through transit observations can constrain the planetary magnetic field, something that is as yet unobserved in exoplanets.

In \citet*[hereafter Paper~2]{paper2}, we applied the shock model initially developed for WASP-12b to the transiting systems known at that point, determining which planets are prone to develop shocks. We predicted that a significant number of transiting systems (36 out of 92 planets) might have a detectable shock, implying that bow shocks might indeed be a common feature surrounding transiting planets. These shocks can cause light curve asymmetries, that once observed can allow us to determine the planetary magnetic field intensity. 

In our previous works \citepalias{paper1,paper2}, we considered these shocks to be static. However, preliminary results from a second set of near-UV observations from Haswell et al. (in prep.) show that the early ingress observed in the near-UV light curve of WASP-12b presents temporal variations. They found that the early ingress was more delayed in the observations of Apr/2010 than in the first set of observations from Sep/2009. This has motivated us to investigate which effects could cause temporal variation in the near-UV light curve of a transiting planet.

Such temporal variation in the light curve of WASP-12b indicates that the stand-off distance between the shock and the planet is varying. According to our model, the shock limits the extension of the magnetosphere of the planet, so that variation in the stand-off distance implies that the size of the planet's magnetosphere is adjusting itself in response to variations in the surrounding ambient medium. 

In this paper we consider possible causes of time-dependence in the location and density of the shocked material around the planet. This will occur if the planet experiences changes in its local environment (such as the density, magnetic field strength and stellar wind speed) as it moves in its orbit. Three causes might be:
\begin{enumerate}
\item The stellar magnetic field is in a steady state and is azimuthally symmetric, but the planet is in an eccentric orbit. Therefore, during the orbit, the distance from the planet to the host star varies. 
\item The stellar magnetic field is in a steady state but is not azimuthally symmetric. During its orbit the planet therefore experiences different local conditions due to these azimuthal magnetic field variations.
\item The stellar magnetic field is intrinsically time-dependent -- either due to turbulent fluctuations in the stellar wind, flaring, coronal mass ejections, or longer-timescale variation due to a magnetic cycle.
\end{enumerate}

Our aim is to investigate how each of these processes may contribute to temporal variation on the planetary bow shock and magnetosphere. In this paper, we consider only systems with high eccentricities, which were neglected in \citetalias{paper2}. 

This paper is organised as follows: Section~\ref{sec.eccentric} discusses the effect on the shock configuration around a planet in an eccentric orbit. Section~\ref{sec.discussion} investigates the three suggested causes of time-dependence in light curve asymmetries. Because the distance from the star changes along the orbit of an eccentric planet, the shock characteristics are modulated by orbital phase. Similarly, non-thermal planetary radio emission should be modulated by orbital phase. We estimate the amplitude of the modulation on the frequency of radio emission in Section~\ref{sec.discussion}. A summary and conclusions are present in Section~\ref{sec.conclusions}.

\section{Bow Shocks Around Eccentric Planets}\label{sec.eccentric} 
In order for a bow shock to form, the condition that needs to be satisfied is that the planetary velocity through the stellar coronal material has to be supersonic. We investigate two different scenarios for the stellar coronal material, which can be (1) static, as in a corona that is magnetically confined to corotate with the host star or (2) dynamic, as a result of the presence of a stellar wind. Shocks can be developed in both scenarios, but the orientation of the bow shock (given by the angle between the normal of the shock and the azimuthal vector -- see Fig.~1 in \citetalias{paper1}) varies depending on the scenario adopted. In scenario (1), the angle is either $\theta=0^{\rm{o}}$, when the shock forms ahead of the planetary orbital motion, or $\theta=180^{\rm{o}}$, when the shock forms behind the planetary orbital motion. In scenario (2), this angle lies between $\theta=0^{\rm{o}}$ and $90^{\rm{o}}$, in general.

The condition that defines whether a shock formed in the magnetically confined case is an ``ahead-shock'' ($\theta=0^{\rm{o}}$) or a ``behind-shock'' ($\theta=180^{\rm{o}}$) depends on whether the coronal material rotates faster than the orbital velocity of the planet. If a prograde planet orbits the star beyond its Keplerian corotation radius, the coronal plasma lags behind the planetary motion, and the shocked material accumulates behind the planetary orbital path, trailing the planet. This gives rise to a behind-shock. CoRoT-11b is a candidate to host a shock in such a configuration. In the opposite case when the planet orbits the star inside its corotation radius, the shock forms ahead of the planetary motion. 

\begin{figure}
	\includegraphics[width=50mm]{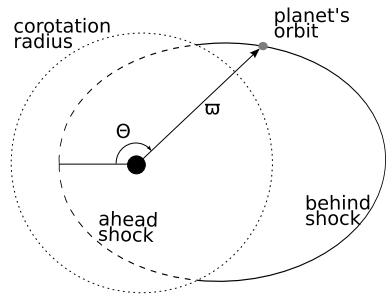}
\caption{Sketch of shock formation. For positions in the orbit where the planet-star distance $\varpi$ is $\varpi<r_{\rm co}$, an ahead-shock is formed. For $\varpi>r_{\rm co}$, a behind-shock exists.}\label{fig.sketch1}
\end{figure}

We note that certain eccentric systems can cross the corotation radius twice during their orbits (Fig.~\ref{fig.sketch1}). The corotation radius $r_{\rm co}$ is given by
\begin{equation}\label{eq.r_co}
\frac{r_{\rm co}}{R_*} = \left( \frac{G M_*}{R_* {[v \sin(i)/\sin(i)]}} \right)^{1/3}.
\end{equation}
During the orbital phases where the distance between the planet and the star ($\varpi$) lies inside $r_{\rm co}$, the shock forms ahead of the planetary motion. For the remaining phases of the orbit ($\varpi>r_{\rm co}$), the shock trails the planet. At the two points in the orbit where $\varpi=r_{\rm co}$, the shock weakens considerably and eventually may disappear, because the net azimuthal velocity between the coronal material and the motion of the planet is small.

Table~\ref{table1} presents the planetary and stellar parameters for the most eccentric transiting planets ($e>0.3$) known to date. The data used here are taken from the exoplanets encyclopedia (http://exoplanet.eu/catalog.php), except for the sky-projected stellar rotation velocity $v \sin(i)$, which was taken from the compilation made by \citet{2010ApJ...719..602S} for most of the systems. For CoRoT-10 and HAT-P-17, values of $v \sin(i)$ were taken from \citet{bonomo2010} and \citet{howard2010}, respectively. Values of impact parameter $b$ and transit duration $t_{\rm dr}$ were compiled from the literature\footnote{For the systems $b$ was not available, we calculated it as $b=(a/R_*) \cos i (1-e^2)/(1+e\sin \omega)$, where $\omega$ is the argument at periastron.} \citep{moutou2009, winn2009, bonomo2010, howard2010, pal2010, queloz2010}. 

\begin{table*}
\caption{Planetary and stellar parameters for the most eccentric transiting planets ($e>0.3$) known to date. The columns are: (1) planet name, (2) planet mass, (3) planet radius, (4) planet orbital period, (5) eccentricity of the orbit, (6) semi-major axis, (7) optical transit duration, (8) impact parameter, distance to the star when the planet is (7)  at periastron and (8) at apastron, (9) stellar mass, (10) stellar radius, (11) sky-projected rotational velocity, (12) stellar corotation radius, and (13) the ratio of the size of planet magnetosphere at apastron and at periastron. \label{table1}}
\begin{center}
\begin{tabular}{lcccccccccccccc}
\hline
Planet&$M_p$&$R_p$&$P_{\rm orb}$&$e$&$a$&$t_{\rm dr}$&$b$&$\varpi_{\rm pe}$&$\varpi_{\rm ap}$&$M_*$&$R_*$&$v \sin(i)$&$r_{\rm co}$& $r_M (\varpi_{\rm ap})/r_M (\varpi_{\rm pe})$\\
Name&$(M_{\rm J})$&$(R_{\rm J})$&(d)&&$(R_*)$&$(h)$&$(R_*)$&$(R_*)$&$(R_*)$&$(M_\odot)$&$(R_\odot)$&(km/s)&$(R_*)$ & \\
\scriptsize{(1)} & \scriptsize{(2)} & \scriptsize{(3)} & \scriptsize{(4)} & \scriptsize{(5)} & \scriptsize{(6)} & \scriptsize{(7)} & \scriptsize{(8)}& \scriptsize{(9)} & \scriptsize{(10)}  & \scriptsize{(11)} & \scriptsize{(12)}& \scriptsize{(13)}& \scriptsize{(14)}& \scriptsize{(15)}\\
\hline \hline
HD 80606b	&	$	3.94	$	&	$	0.92	$	&	$	111.44	$	&	$	0.93	$	&	$	99	$	&	$	9.5-17.2	$	&	$	0.75	$	&	$	7	$	&	$	191	$	&	$	0.90	$	&	$	0.98	$	&	$	1.8	$	&	$	39	$	&	$	29.1	$	\\
HD 17156b	&	$	3.21	$	&	$	1.02	$	&	$	21.22	$	&	$	0.68	$	&	$	24	$	&	$	3.177	$	&	$	0.55	$	&	$	8	$	&	$	40	$	&	$	1.24	$	&	$	1.45	$	&	$	2.6	$	&	$	27	$	&	$	5.2	$	\\
CoRoT-10b	&	$	2.75	$	&	$	0.97	$	&	$	13.24	$	&	$	0.53	$	&	$	29	$	&	$	2.98	$	&	$	0.85	$	&	$	13	$	&	$	44	$	&	$	0.89	$	&	$	0.79	$	&	$	2	$	&	$	39	$	&	$	3.3	$	\\
HAT-P-2b	&	$	8.74	$	&	$	1.19	$	&	$	5.63	$	&	$	0.52	$	&	$	9	$	&	$	4.29	$	&	$	0.395	$	&	$	4	$	&	$	13	$	&	$	1.36	$	&	$	1.64	$	&	$	20.8	$	&	$	6	$	&	$	3.1	$	\\
HAT-P-17b	&	$	0.53	$	&	$	1.01	$	&	$	10.34	$	&	$	0.35	$	&	$	23	$	&	$	4.06	$	&	$	0.311	$	&	$	15	$	&	$	30	$	&	$	0.86	$	&	$	0.84	$	&	$	0.3	$	&	$	136	$	&	$	2.1	$	\\
WASP-8b	&	$	2.24	$	&	$	1.04	$	&	$	8.16	$	&	$	0.31	$	&	$	18	$	&	$	4.40	$	&	$	0.604	$	&	$	12	$	&	$	24	$	&	$	1.03	$	&	$	0.95	$	&	$	2	$	&	$	37	$	&	$	1.9	$	\\
\hline
\end{tabular}
\end{center}
\end{table*}

For an elliptical orbit, the distance $\varpi$ from the planet to the star (located at one of the focus of the ellipse) is given by
\begin{equation}\label{eq.varpi}
\varpi = \frac{a (1-e^2)}{1+e \cos(\Theta)}
\end{equation}
where $e$ is the eccentricity and $\Theta$ is the angle formed between the connecting planet-star line at nearest approach (periastron) and the position of the planet. Therefore, at periastron, $\Theta = 0^{\rm o}$, while at apastron $\Theta = 180^{\rm o}$. The radial and tangential velocities of the planet (assuming $M_* \gg M_p$) are, respectively,
\begin{equation}\label{eq.vpl-r}
v_{r, pl} = \left( \frac{G M_*}{a (1-e^2)}\right)^{1/2} e \sin(\Theta)
\end{equation}
and
\begin{equation}\label{eq.vpl-phi}
v_{\varphi, pl} = \left( \frac{G M_*}{a (1-e^2)} \right)^{1/2} [1+ e \cos(\Theta)].
\end{equation}

The condition for a shock to be formed is that the relative velocity between the material surrounding the planet and the planet itself $|\Delta {\bf u}|$ must exceed the sound speed $c_s=(k_B T/m)^{1/2}$, where $k_B$ is the Boltzmann constant, $m=\mu m_p$ is the mean particle mass, with $\mu =0.66$ and $m_p$ the proton mass. Therefore, there exists a maximum temperature below which shock formation is allowed
\begin{equation}\label{eq.tmax}
T_{\rm max} = \frac{|\Delta {\bf u}|^2 m}{k_B} .
\end{equation}

Here, we make a distinction between two possible scenarios: (1) a corona that is magnetically confined to corotate with the host star and (2) a case where the coronal material expands in the form of a stellar wind. In the magnetically confined scenario, the coronal material has only rotation (azimuthal) velocity given by 
\begin{equation}\label{eq.u_cor}
u_{\varphi , \rm{cor}} = {\Omega_*}{\varpi} = \frac{v \sin(i)}{\sin(i)} \frac{\varpi}{R_*},
\end{equation}
where the angular velocity of the star $\Omega_* = [v \sin(i)/\sin(i)]/ R_*$. In the wind scenario, the material surrounding the planet has both a rotation velocity (conservation of angular momentum of particles in the wind)
\begin{equation}
u_{\varphi , w} = \frac{\Omega_* R_*^2}{\varpi} = \frac{v \sin(i)}{\sin(i)} \frac{R_*}{\varpi}
\end{equation}
and a radial velocity $u_{r,w}$ due to the stellar wind. By assuming that the host star has an isothermal, thermally-driven wind \citep{parker}, we calculate the wind velocity profile from the integration of the differential equation
\begin{equation}  \label{eq.wind-vel}
n_w m u_{r,w} \frac{\partial u_{r,w}}{\partial r} = -\frac{\partial p_w}{\partial r} - n_w m \frac{G M_*}{r^2} , 
\end{equation} 
where $n_w$ is the wind density, $r$ is the radial distance from the star, $p_w = n_w k_B T$ is the wind thermal pressure.

Therefore, the relative velocity $|\Delta {\bf u}|$ between the material surrounding the planet and the planet itself for the confined case is
\begin{equation}\label{eq.deltau-cor}
|\Delta {\bf u}| = |{\bf u}_{{\rm cor}} - {\bf v}_{pl}| =  [v_{r, pl}^2 + (u_{\varphi , {\rm cor}} - u_{\varphi, pl})^2 ]^{1/2}
\end{equation}
and for the wind case is
\begin{equation}\label{eq.deltau-w}
|\Delta {\bf u}| = |{\bf u}_{\rm w} - {\bf v}_{pl}| = [(u_{r,w}-v_{r, pl})^2 + (u_{\varphi , w} - u_{\varphi, pl})^2]^{1/2}.
\end{equation}
%

\subsection{HAT-P-2b}\label{sec.hatp2}
\begin{figure}
	\includegraphics[width=84mm]{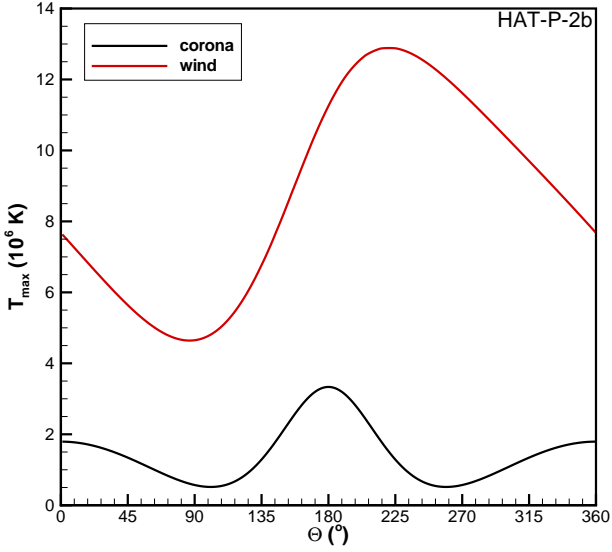}
\caption{Maximum temperature required for shock formation for HAT-P-2b for both the confined corona scenario (black) and the wind scenario (red). At $\Theta=0, 360^{\rm o}$, the planet is at periastron and at $\Theta=180^{\rm o}$ at apastron. At orbital phases corresponding to $\Theta \simeq 101^{\rm o}$ and $\Theta \simeq 259^{\rm o}$, the net velocity between the planet and the coronal material is very small, so the temperature required for shock formation in the confined corona scenario should also be very small, implying that a shock will not be formed near these orbital phases. Considering $T=2 \times 10^6$~K for the wind scenario.}\label{fig.Tmax-hatp2}
\end{figure}

HAT-P-2b is a good illustrative example to compare both scenarios for the surrounding stellar material, since its orbit may lie almost entirely within the stellar magnetosphere and, in this case, the hypothesis adopted in the confined scenario case should be met during almost its entire orbit. From equations~(\ref{eq.tmax}), (\ref{eq.deltau-cor}) and (\ref{eq.deltau-w}), we show in Fig.~\ref{fig.Tmax-hatp2} the maximum temperature required for shock formation for HAT-P-2b for both the confined corona scenario (black) and the wind scenario (red), where we adopted a coronal temperature of $T=2\times 10^6$~K in the evaluation of $u_{r,w}$ (equation~\ref{eq.wind-vel}). It is interesting to note that at orbital phases near $\Theta\simeq 101^{\rm o}$ and $\Theta\simeq 259^{\rm o}$, the velocity difference of the confined corona case approaches its minimum and a shock will only be formed if the temperature is very small ($T < T_{\rm max} \simeq 5 \times 10^5$~K). Because such a small temperature is not expected in the corona of a F8-star, it is probable that a shock will not be formed near these orbital phases. Furthermore, for the confined corona scenario, if the coronal temperature of HAT-P-2b is $ T = 2\times 10^6$~K, a shock will not be formed for orbital phases where $T_{\rm max} > T = 2\times 10^6$~K, i.e., between $\Theta \simeq 158^{\rm o}$ and $212^{\rm o}$ (around apastron).

\begin{figure}
	\includegraphics[width=84mm]{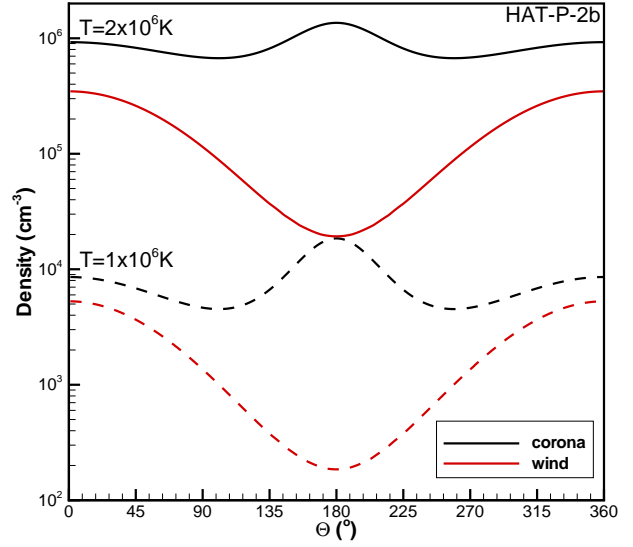}
\caption{Density of the ambient medium through where the planet HAT-P-2b orbits as a function of planetary orbital phase (given by $\Theta$). Two different temperatures are assumed: $T=1 \times 10^6$~K (dashed lines) and $T=2 \times 10^6$~K (solid lines), for both the confined corona (black lines) and the wind (red lines) scenarios.}\label{fig.density-hatp2}
\end{figure}

We note that even if a shock is formed around the planet, it may not be detected if the compressed local plasma does not achieve a density high enough to cause an observable level of optical depth. For the corona case, the local density of the corona is given by
\begin{equation}\label{eq.dens-hyd}
 n_{\rm cor} = n_0\exp\left[ \frac{GM_*/R_*}{k_B T/m} \left( \frac{R_*}{R_{\rm orb}} -1 \right) + \frac{[v \sin(i)]^2}{2 k_B T/m} \left( \frac{R_{\rm orb}^2}{R_*^2}-1 \right)\right] ,
\end{equation}
where $n_0$ is the density at the base of the corona. For simplicity, we assume that $n_0=10^8$~cm$^{-3}$, similar to that of the solar corona \citep{1988ApJ...325..442W}. For the wind case, the density $n_w$ is given by conservation of mass of the wind $n_w u_{r,w} r^2 = {\rm constant}$, with $u_{r,w}$ given by equation~(\ref{eq.wind-vel}). Fig.~\ref{fig.density-hatp2} shows how the density varies according to the planetary orbital phase (given by $\Theta$) for HAT-P-2b assuming two different temperatures for the material that surrounds the planet. Because HAT-P-2 has a relatively high $v \sin (i)$, the density at apastron is larger than at periastron in the confined case (the second term in the RHS of equation~(\ref{eq.dens-hyd}) dominates over the first term). We note, however, that these `bumps' may not be physical, since the corona may not still be corotating with the star at large distances. A way to estimate where this assumption breaks down is to evaluate the plasma-$\beta$, the ratio between thermal and magnetic energy densities: whenever the thermal pressure takes over the plasma pressure, the hypothesis may become invalidated. 

In \citetalias{paper2}, we suggested that a local density surrounding the planet as low as $10^4$~cm$^{-3}$ can still provide detection. Therefore, according to Fig.~\ref{fig.density-hatp2} detection of shocks around HAT-P-2b implies that the coronal temperature should be $T \gtrsim 2\times 10^6$~K. Note, however, that because a shock does not exist during orbital phases $158^{\rm o} \lesssim \Theta \lesssim 212^{\rm o}$ (when $T>T_{\rm max}$, assuming $T=2\times 10^6$~K), shock formation and detection around a planet can constrain both its surrounding density and temperature.

\begin{figure}
	\includegraphics[width=84mm]{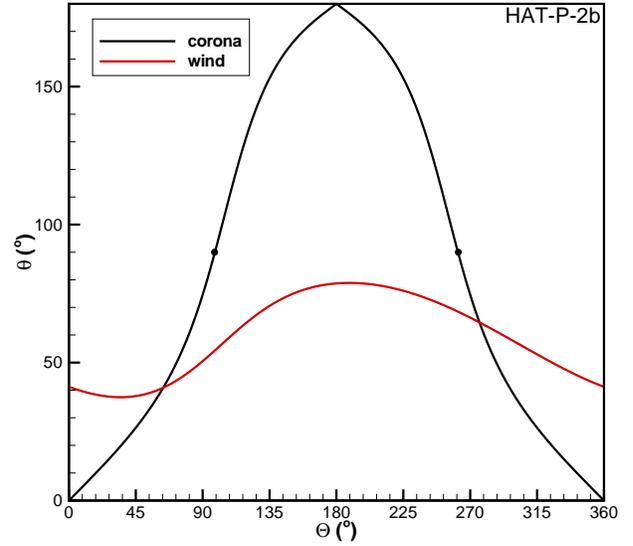}
\caption{The angle $\theta$ that the shock normal makes to the relative azimuthal velocity of the planet as a function of orbital phase given by $\Theta$, for HAT-P-2b. A purely ahead-shock ($\theta =0^{\rm o}$) occurs at periastron ($\Theta = 0, 360^{\rm o}$). A purely behind-shock ($\theta =180^{\rm o}$) occurs at apastron ($\Theta = 180^{\rm o}$). A day-side shock ($\theta =90^{\rm o}$) occurs at $\Theta = 98, 262^{\rm o}$ (positions marked with filled circles).}\label{fig.angle-hatp2}
\end{figure}

We define the angle that the shock normal makes to the relative azimuthal velocity of the planet as
\begin{equation}\label{eq.angle-cor}
\theta =  \arctan{\left(\frac{|v_{r, pl}|}{v_{\varphi, pl} - u_{\varphi, cor}}\right) } ,
\end{equation}
for the  confined corona scenario and for the wind scenario as
\begin{equation}\label{eq.angle-wind}
\theta =  \arctan{\left(\frac{|v_{r, pl} - u_{r, w} |}{v_{\varphi, pl}- u_{\varphi, w} }\right) } .
\end{equation}
Due to the $\Theta$-dependence of the involved velocities, the angle of the shock $\theta$ also depends on the orbital phase. Fig.~\ref{fig.angle-hatp2} shows the shock angle $\theta$ for both the corona (black) and wind (red) scenarios for HAT-P-2b assuming $T=2\times 10^6$~K. For eccentric systems in the confined scenario, we note that a purely ahead-shock ($\theta =0^{\rm o}$) or behind-shock ($\theta =180^{\rm o}$) only occurs when the radial component of the orbital velocity is null ($v_{r, pl} = 0$ in equation \ref{eq.angle-cor}), which happens at periastron ($\Theta = 0, 360^{\rm o}$) and apastron ($\Theta = 180^{\rm o}$), respectively. A purely day-side shock ($\theta =90^{\rm o}$) happens when $u_{\varphi, cor} - v_{\varphi, pl} =0$, which occurs at orbital phases $\Theta = 98, 262^{\rm o}$, indicated by filled circles in Fig.~\ref{fig.angle-hatp2}. For the wind case, because $u_{\varphi, w} < v_{\varphi, pl}$ in general, $\theta< 90^{\rm o}$, implying that the shock never forms behind the planetary motion. For higher wind temperatures, we expect the shock angle to approach $90^{\rm o}$ \citepalias{paper1}. 

The extent of the planetary magnetosphere $r_M$ can be determined by static pressure balance\footnote{We are correcting a typographical error in equation (9) of \citetalias{paper1} and equation (5) of \citetalias{paper2}. This has no effect on the calculations done in these papers.}
\begin{equation}\label{eq.equilibrium}
{m n \Delta u^2} + \frac{B (\varpi)^2}{8\pi} + p= \frac{B_{p}(r_M)^2}{8\pi} + p_{p} .
\end{equation}
Both $n$ and $p$ are the local density and pressure of the material surrounding the planet, which can either refer to the coronal scenario ($n_{\rm cor}$, $p_{\rm cor} = n_{\rm cor} k_B T$) or to the wind scenario ($n_w$, $p_w$). $B (\varpi)$ is the stellar magnetic field intensity at the site of the interaction with the planet. In the confined case, we assume that the stellar magnetic field is dipolar so that $B (\varpi) = {B_*}(R_*/\varpi)^{3}$. In the case where the corona escapes in the form of a wind, we neglect the effect of the stellar magnetic field. We also consider that the planet magnetic field geometry can be approximated as that of a dipolar field, so $ B_{p}(r_M) = {B_p}{(R_p/r_M )^{3}}$, where $B_p$ and $B_{p}(r_M)$ are the magnetic field strengths at the surface of the planet and at the nose of the planet's magnetosphere, respectively. We neglect the planet thermal pressure $p_p$. 

If the dominant terms in equation~(\ref{eq.equilibrium}) are the magnetic terms, we showed that, for the confined corona scenario \citepalias{paper1}, equation~(\ref{eq.equilibrium}) reduces to
\begin{equation}
\frac{r_M }{R_p} =\left(\frac{B_p}{B_*} \right)^{1/3} \frac{\varpi}{R_*}.
\end{equation}
In this case, the ratio between $r_M$ at apastron and at periastron is 
\begin{equation}
\frac{r_M (\varpi_{\rm ap})}{r_M (\varpi_{\rm pe})} = \frac{1+e}{1-e},
\end{equation}
where we used equation~(\ref{eq.varpi}). For HAT-P-2b, ${r_M (\varpi_{\rm ap})}/{r_M (\varpi_{\rm pe})} \sim 3.1$. Table~\ref{table1}, column 13 presents this value for the remaining planets.

In our previous works \citepalias{paper1, paper2}, for the confined corona case with a planet in a circular orbit, we assumed that the observed stand-off distance of the shock from the planetary surface is approximately the extent of the planetary magnetosphere $r_M$. The stand-off distance can be observed through the time difference of the ingress of two transit light curves taken at different wavelengths \citepalias{paper1}. Therefore, it is straightforward to derive the size of the magnetosphere of the planet and the ratio $B_p/B_*$ for circular orbits in a confined corona. If the magnetic field of the star is known, then the planetary magnetic field is easily derived. For the other cases (i.e., in the wind scenario and/or for eccentric systems), the shock is not always formed purely ahead of the planet, as shown in Fig.~\ref{fig.angle-hatp2}. Therefore, from observations, one can derived the stand-off distance of the shock that is {\it projected} on the plane of the sky. 

For the calculations presented hereafter, we consider all the terms in equation~(\ref{eq.equilibrium}), except $p_p$. Detailed calculation of pressure balance as a means to determine the size of a planet's magnetosphere has been used by the Solar System community for several decades \citep{1930Natur.126..129C}. In these calculations, a correcting factor is often assumed in the value of $B_{p}(r_M)$ (equation \ref{eq.equilibrium}) to account for the compression of the magnetic field at the nose of the magnetosphere. In the case of a dipolar planetary magnetic field, the presence of magnetopause currents gives rise to a magnetic field with the same intensity as the magnetic field just inside the magnetosphere \citep[e.g., ][]{1997pssp.conf.....C, 2007bsw..book.....M}, doubling, therefore, the value of $B_{p}(r_M)$. To allow for different magnetopause geometries, it is also considered that the magnetic field at the nose of the magnetopause increases by a factor $f$ \citep[e.g., ][]{2004A&A...425..753G, Khodachenko2007}. Altogether, these modifications result that $B_{p}(r_M) \to 2 f B_{p}(r_M)$ in equation~(\ref{eq.equilibrium}). For the Earth's magnetosphere, the compression factor is empirically determined as $2f \simeq 2.44$ \citep{1995isp..book.....K}. Given the other uncertainties involved in the calculation of $r_M$ for exoplanets, we do not include the compression factor $2f$ in this work. If it were considered, the magnetospheric radii derived in this paper would have increased by a factor of $(2 f)^{1/3}$.
 
\begin{figure}
	\includegraphics[width=84mm]{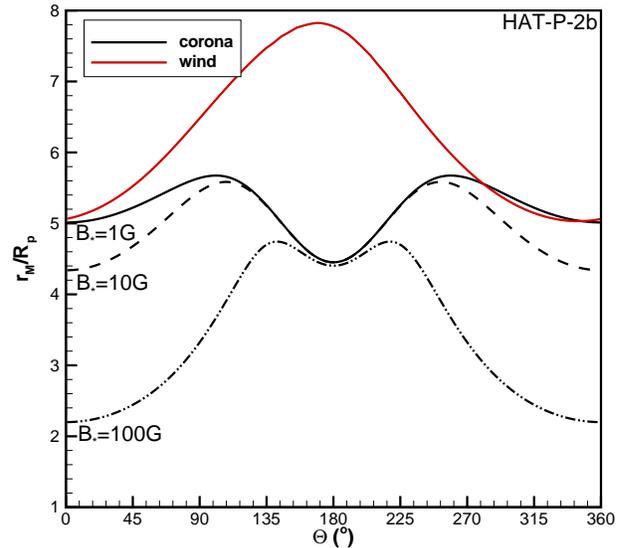}
\caption{The magnetospheric size $r_M$ of HAT-P-2b as a function of planetary orbital phase given by $\Theta$ assuming a coronal temperature $T = 2 \times 10^6$~K and $B_p = 14$~G. Black curves refer to the confined corona scenario with different $B_*$ and the wind scenario is represented by the red solid line.}\label{fig.magnetosphere}
\end{figure}

Fig.~\ref{fig.magnetosphere} shows the magnetospheric radius of HAT-P-2b derived using equation~(\ref{eq.equilibrium}) for both the wind case (red) and the confined corona case (black) for $B_p = 14$~G (similar to that of Jupiter) and $B_*=1$~G (solid), $10$~G (dashed) and $100$~G (dot-dashed). For the confined case, we note that the more intense $B_*$ is, the more compact is the magnetosphere of HAT-P-2b. The magnetospheric sizes of the Earth, Jupiter, Saturn, Uranus and Neptune are $\simeq 11$, $50-100$, $16-22$, $18$, and $23-26$ planetary radii, respectively \citep{bagenal1992}. These values are larger than the values derived for HAT-P-2b (Fig.~\ref{fig.magnetosphere}). The compact size of the magnetosphere of HAT-P-2b is because the planet is located much closer to the star where the stellar wind pressures are still significant.

\begin{figure}
	\includegraphics[width=84mm]{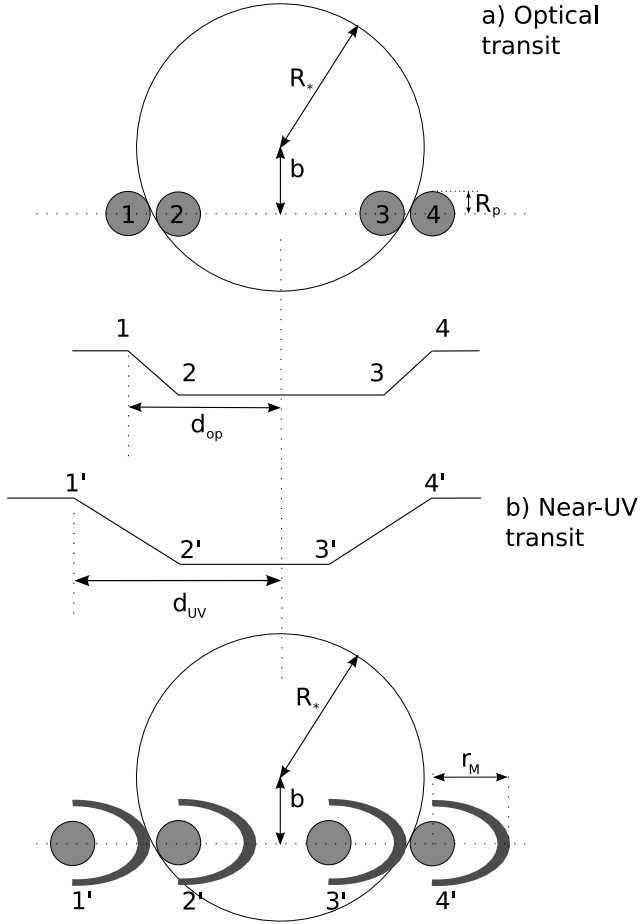}
\caption{Sketches of the light curves obtained through observations in the a) optical and b) near-UV, where the bow shock surrounding the planet's magnetosphere is also able to absorb stellar radiation.}\label{fig.sketch2}
\end{figure}

\begin{figure}
	\includegraphics[width=84mm]{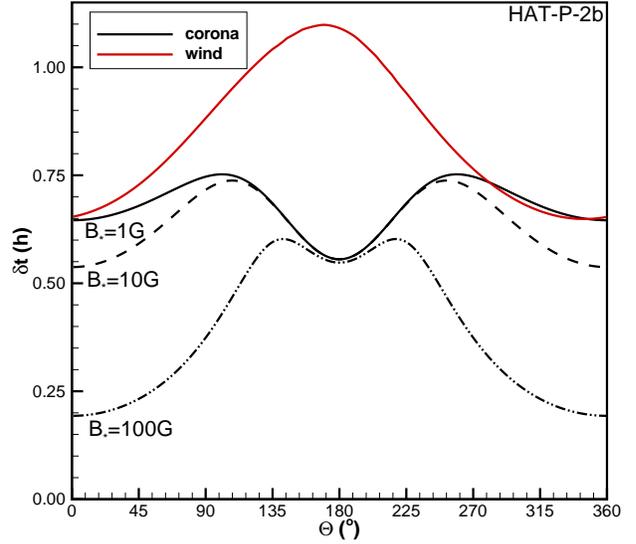}
\caption{Expected time differences $\delta t$ between the beginnings of the optical and near-UV transits calculated for HAT-P-2b. Curve labels are as in Fig.~\ref{fig.magnetosphere}.}\label{fig.delta-t}
\end{figure}

The time difference $\delta t$ between the beginning of the optical and near-UV transits led us to suggest the existence of a bow shock around the hot-Jupiter WASP-12b \citepalias{paper1}. This time difference can be inferred by geometrical considerations of the transiting systems. Consider the sketches presented in Fig.~\ref{fig.sketch2}. $d_{\rm op}$ and $d_{\rm UV}$ are, respectively, the sky-projected distances that the planet (optical) and the system planet+magnetosphere (near-UV) travel from the beginning of the transit until the middle of the optical transit
\begin{equation}
d_{\rm op} = (R_*^2 - b^2)^{1/2} + R_p
\end{equation}
and
\begin{equation}
d_{\rm UV} = (R_*^2 - b^2)^{1/2} + r_M,
\end{equation}
where $b$ is the impact parameter derived from transit observations. The optical transit duration $t_{\rm dr}$ is the time the planet takes to travel $2 d_{\rm op}$ (from beginning to end of the optical transit). Therefore, by a linear extrapolation we derive
\begin{equation}
\delta t = \frac{t_{\rm dr}}{2} \left( \frac{d_{\rm UV}}{d_{\rm op}} - 1\right).
\end{equation}
Adopting $b=0.395~R_*$ and $t_{\rm dr} = 0.1787$~d \citep{pal2010}, Fig.~\ref{fig.delta-t} shows the time-offset $\delta t$ for HAT-P-2b for the confined corona (black) and wind (red) scenarios. We adopt $B_p = 14$~G and $B_*=1$~G (solid), $10$~G (dashed) and $100$~G (dot-dashed). We again note the modulation of $\delta t$ with planetary orbital phase. For $B_* \gtrsim 10$~G, $\delta t$ becomes comparable do the time offset observed for WASP-12b ($\delta t \simeq 0.5~$h).

\subsection{Remaining Eccentric Systems}
For a shock to exist and be observable, the following conditions must be met: 1) the temperature of the ambient medium surrounding the planet has to be $T<T_{\rm max}$ (equation~\ref{eq.tmax}); 2) the local density of this ambient has to be larger than a minimum threshold to provide enough detectable absorption ($n \gtrsim 10^4~$cm$^{-3}$); 3) the magnetosphere of the planet has to be large enough to provide a time offset that is larger than the instrumental uncertainties in time.

\begin{figure}
	\includegraphics[width=84mm]{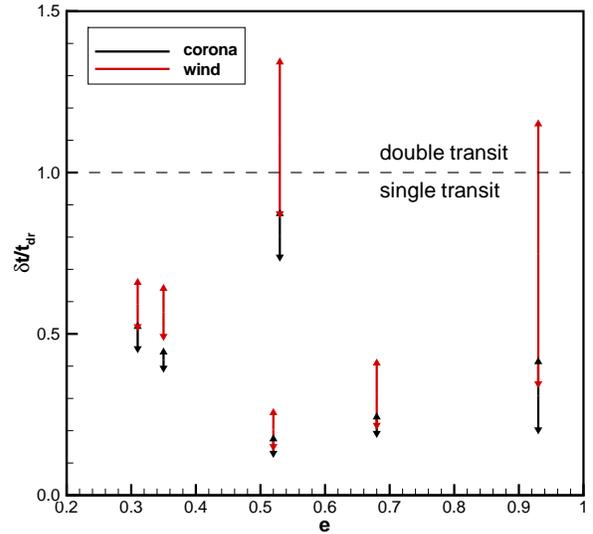}
\caption{Time-delay $\delta t$ between optical and near-UV transit beginnings scaled to the transit duration $t_{\rm dr}$ for the eccentric systems studied in this paper (Table~\ref{table1}). Planets are shown in order of increasing eccentricity: WASP-8b, HAT-P-17b, HAT-P-2b, CoRoT-10b, HD17156b, HD80606b.}\label{fig.delta-t-all}
\end{figure}

In order to evaluate the time differences $\delta t$ between the beginnings of the optical and near-UV transits for the remaining systems presented in Table~\ref{table1}, we performed a similar analysis to that presented in Section~\ref{sec.hatp2}. We assume that all host stars analysed in this paper have coronal winds with $T = 2\times 10^6$~K and base densities $n_0=10^8$~cm$^{-3}$ and planets have $B_p=14$~G. We evaluated $\delta t$ for both the confined corona and wind cases. Fig.~\ref{fig.delta-t-all} shows the minimum and maximum expected time offset $\delta t$ scaled to the transit duration $t_{\rm dr}$ for a stellar magnetic field of $B_*=1$~G. The confined corona scenario (black) provides smaller time offsets than the ones derived in the wind scenario (red). In Fig.~\ref{fig.delta-t-all}, we assume that shocks can be formed at all points in the planetary orbit, although this may not always be the case (see previous paragraph).

Near-UV light curve asymmetry has been detected in the giant planet WASP-12b \citep{fossati2010}. So far, we do not know the precise location of the material that is causing the absorption observed in Mg~II lines. It may be that absorption is concentrated in a certain location of the planetary atmosphere (e.g., only around the shocked material) or that it extends through a larger area of the planetary atmosphere. Observations of near-UV light curves from different systems can help constrain the location of absorption. 

\begin{figure*}
	\includegraphics[width=150mm]{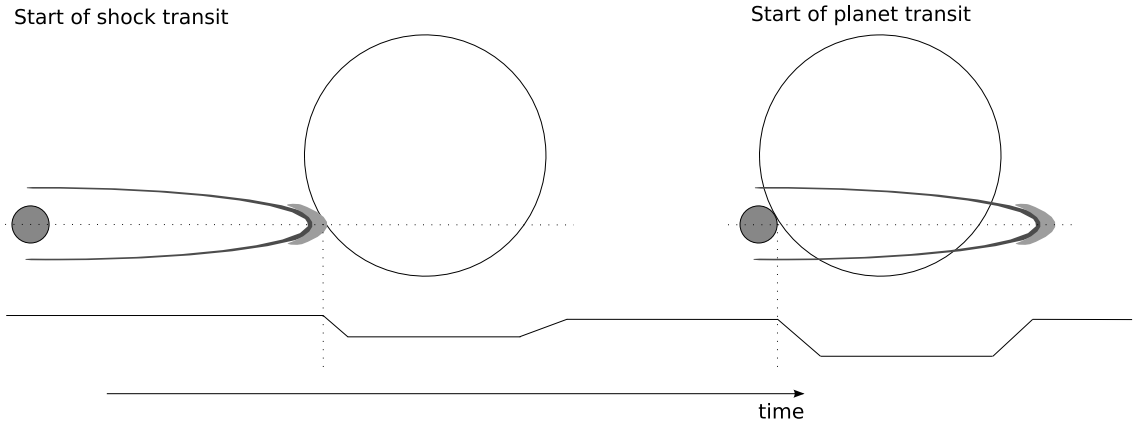}
\caption{Sketch of the double transit caused by a very extended planetary magnetosphere. If the material absorbing near-UV stellar radiation is concentrated in a small region around the bow shock (see shaded area), it ingresses in transit before the planetary transit. This causes the first dip in the light curve. After its transit, the light curve returns to its continuum level until the planet itself ingresses in the transit (second dip). If the absorbing region fills the magnetosphere of the planet, the transit of the material and the planet are blended together and only one transit is observed.}\label{fig.sketch3}
\end{figure*}

Consider the case where the material causing absorption in Mg~II lines is restricted around the small region of the shocked material, which is located at a distance $\sim r_M$ from the center of the planet. If this distance is larger than the path length of the transit ($2[R_*^2-b^2]^{1/2}$), two independent transits occurs (Fig.~\ref{fig.sketch3}). The first one is the transit of the shocked material itself, which ends before the planetary transit begins. After the first transit, the light curve returns to the continuum level until the planetary transit itself takes place. We note that if one just observes immediately before the planetary transit, the transit of the shocked material may be missed. Now, consider that the material causing absorption in Mg~II lines fills the magnetosphere of the planet, still assuming that $r_M>2(R_*^2-b^2)^{1/2}$. Instead of causing a double transit, the transit of the shocked material will blend with the transit of the planet, and only one transit occurs.

The condition $r_M>2(R_*^2-b^2)^{1/2}$ can be written in terms of the transit offset and duration as $\delta t/t_{\rm dr}>1$ ($\delta t/t_{\rm dr}=1$: dashed line in Fig.~\ref{fig.delta-t-all}). Among the eccentric systems analysed in this paper with the assumed values of $T_0=2\times 10^6$~K, $B_p=14$~G, $B_*=1$~G and $n_0=10^8$~cm$^{-3}$, we note that there are only two systems suitable to probe the extension of the absorbing material: HD~80606b and CoRoT10b. These systems could present a double transit if absorption is restricted to a small area surrounding the shock  or a single transit if absorption occurs at an extended area. We note that the largest values of time offsets in Fig.~\ref{fig.delta-t-all} happen during apastron, when the size of the magnetosphere is the largest one (wind case). However, during apastron the material surrounding the planet has its smallest density. Therefore, it is more difficult to observed shock at apastron, where double transits are more probable to happen.

The case where the magnetosphere of the planet is contained inside the projected distance (i.e., $r_M<(R_*^2-b^2)^{1/2}$ or, equivalently, $\delta t < t_{\rm dr}$) always results in a single transit. 

\section{Discussion}\label{sec.discussion}

Bow shocks around planets may not be steady. If, for instance, a planet has a thermal mass loss that intrinsically varies, the bow shock formed at the edge of the planetary magnetosphere could expand whenever there is any increase in the planetary mass-loss rate \citep[for mass-loss variations due to the evolution of the host star, see][]{murray2009}. Also, if the dynamo responsible for generating the planetary magnetic field changes its efficiency, the size of the planet's magnetosphere should change accordingly. 

Another circumstance that may cause the position and density of the shocked material around the planet to vary is related to variations not in the planet itself, but in the ambient medium that surrounds it. The last alternative is the focus of our paper. Here, we discuss three possible causes that could lead to such variations.

\subsection{Case 1: Eccentric Planet}
The first case we discuss is the case where the confined corona or coronal wind is in a steady-state and is axisymmetric. We also assume that the orbital axis of the planet is aligned with the stellar axis of symmetry. If the planet is in a circular orbit, during its revolution around the star, the stellar material encountered by the planet does not change its properties. Therefore, in the reference frame of the planet, the shock is steady. However, an eccentric planet probes different plasma conditions as its distance to the host star varies through its orbit. This case was detailed in Section~\ref{sec.eccentric}, where we showed that the characteristics of the bow shock formed around the planet is modulated by the orbital phase. We also showed that at certain phases, a shock may not be formed (depending on the temperature of the coronal material) and that even if a shock is formed, it may not have enough density to be detected.

This variation in shock characteristics can only be probed if the shock is observed at different orbital phases, which is not the case for transit observations that are, in general, conducted at only one orbital phase. However, any radio emission produced by the interaction of the star and planet should be modulated by orbital phase. The amplitude of such modulation is estimated next.

\subsubsection{Planetary Radio Emission}\label{sec.radio}
The four giant planets of the Solar System and the Earth emit at radio wavelengths. Because there are correlations between this emission and the local characteristics of the solar wind, planetary radio emission is associated to the interaction of the solar wind with the planets. In analogy to the Solar System, it has been suggested that exoplanets may also produce radio emission when they interact with the winds of their host stars \citep{jardine2008}. 

\begin{figure}
\includegraphics[width=50mm]{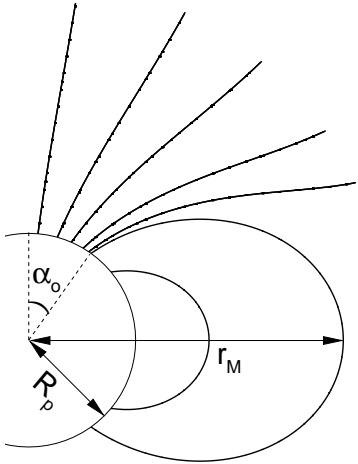}
\caption{Sketch of the planet's magnetic field lines. The closed-field line region extends out to $r_M$. The area where magnetic field lines are open and exposed to the stellar wind (auroral region) extends out to planetary colatitudes $\alpha_0$, which is determined by the magnetosphere's size $r_M$ (equation \ref{eq.alpha0}). Radio emission is believed to arise from a ring surrounding the auroral region.}\label{fig.sketch4}
\end{figure}

In Jupiter, decametric emission is thought to arise in a ring surrounding the auroral region wherein the planetary magnetic field lines are open. The aperture of the auroral ring can be related to the size of the planet's magnetosphere $r_M$ in the following way. We consider the planet to have a dipolar magnetic field aligned with the planetary orbital spin axis, such that 
\begin{equation}
B_{p,r} = B_p  \left (\frac{R_p}{R} \right )^{3} \cos \alpha
\end{equation}
 and 
\begin{equation}
B_{p,\alpha} = \frac{B_p}{2} \left(\frac{R_p}{R} \right )^{3} \sin \alpha , 
\end{equation}
where $R$ is the radial distance from the planet, $B_p$ is the planetary magnetic field evaluated at the pole and $\alpha$ is the angle that measures the colatitude (Fig.~\ref{fig.sketch4}). The paths of individual field lines are described by $A = \sin^2 \alpha / R$ where the value of $A$ labels each field line (i.e., it is constant for each line). If the largest closed field line connects the surface at a colatitude $\alpha_0$ (where $A=\sin^2 \alpha_0/R_p$) to an equatorial radius $R=r_M$ (where $A=1/r_M$), then 
\begin{equation}
\frac{1}{r_M} = \frac{\sin^2 \alpha_0}{R_p} .
\end{equation}
The fractional area of the planetary surface that has open magnetic field lines is then $(1-\cos \alpha_0)$, where we considered both the north and south auroral caps of the planet. Thus, for example, if $r_M=5~R_p$, then $\alpha_0 \simeq 27^{\rm o}$ and open field covers $\sim 11\%$ of the surface. If $r_M=2~R_p$, then $\alpha_0\simeq 45^{\rm o}$ and open field covers $\sim 29 \%$ of the surface. If the planetary radio emission comes from the region that separates closed and open field lines (auroral ring), such emission would happen in a cone with aperture given by
\begin{equation}\label{eq.alpha0}
\alpha_0 = \arcsin{[ (R_p/r_M)^{1/2}} ].
\end{equation}
The planetary magnetic field at this colatitude $\alpha_0$ is
\begin{equation}
B(\alpha_0) = \frac{B_p}{2} (1+3\cos^2 \alpha_0)^{1/2},
\end{equation}
which corresponds to an electron cyclotron frequency
\begin{equation}
f_c =  2.8 \left( \frac{B(\alpha_0)}{1~{\rm G}}\right) ~ {\rm MHz}.
\end{equation}

\begin{figure}
	\includegraphics[width=84mm]{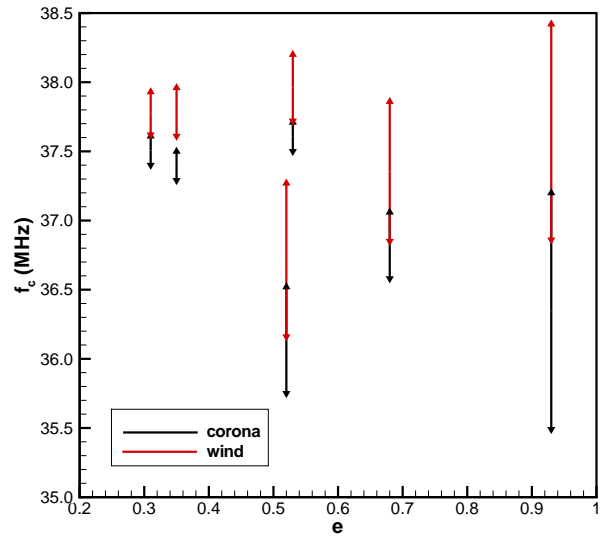}
\caption{Cyclotron frequency $f_c$ of the electrons believed to be responsible for generating planetary radio emission. The bandwidth where radio emission should occurs believed to be a fraction of $f_c$. Planets are shown in order of increasing eccentricity: WASP-8b, HAT-P-17b, HAT-P-2b, CoRoT-10b, HD17156b, HD80606b.}\label{fig.fcyclotron}
\end{figure}

Despite several searches, radio emission from exoplanets has not yet been detected. One possible reason for that may be that emission occurs at a frequency much different from the frequency range of the radio detectors used in the observations. The frequency of the radio emission is believed to be a fraction of $f_c$. As the planetary magnetospheric size is modulated by its orbital phase in an eccentric system (Section~\ref{sec.eccentric}), the latitude of the last closed field line and, consequently, the fraction of the area of the planet that will be exposed to the stellar wind are also modulated by the orbital phase. Therefore, we expect that $f_c$ will present a similar modulation and the frequency where radio emission occurs should differ at different parts of the orbit. In Fig.~\ref{fig.fcyclotron}, we present the minimum and maximum cyclotron frequencies for each of the systems presented in Table~\ref{table1} for the case $B_p=14~$G and $B_*=1$~G. We note from Fig.~\ref{fig.fcyclotron} that $f_c$ is larger for the wind scenario than for the confined case. This is because the size of the planetary magnetosphere is smaller for the latter one. We also note that the modulation in $f_c$ does not vary significantly with planetary orbital phase, reaching an amplitude of only a few MHz and a frequency centred at around $37$~MHz. Although small, this modulation could be spectrally resolved with radio telescopes, such as LOFAR.

Because radio emission is related to the power of the impacting coronal material on the planet, it is preferable to look for radio emission during the closest approach of the planet. Radio emission from HD80606b during the periastron was recently investigated by \citet{2010AJ....140.1929L}, although the detection was unsuccessful. Because the star-planet distance during periastron for the HAT-P-2b is about half the value of the same distance for HD80606b, HAT-P-2b should be a potential system to look for radio emission.

\subsection{Case 2: Azimuthal variations in the coronal material}
The second case we consider is the case where the coronal material is also in a steady state in the stellar reference frame, but it is non-axisymmetric. Because the stellar rotation is not in general locked to the planetary orbital period (except for the case the planet is in the corotation radius of the star), as the planet revolves around the star, it encounters material with different characteristics, causing shock variations. By performing several observations of transit, one can infer the impact on the planetary magnetosphere and bow shock of azimuthal variations in the stellar coronal material. 

An example of a non-axisymmetric coronal material is the case of an oblique stellar magnetosphere, where the stellar spin axis is not parallel to the magnetic moment vector \citep{Vidotto2010}. During its orbit, the planet may encounter a region of confined coronal material (e.g., when it is inside the closed magnetic loops of the host star), but at other orbital phases, it may be in the wind of the host star (where the magnetic field lines of the star are open and the stellar wind escapes). As the characteristics of the closed corona are different to the open-wind region as demonstrated in Section~\ref{sec.eccentric} (Figs.~\ref{fig.Tmax-hatp2} -- \ref{fig.delta-t}), the magnetospheric size of the planet varies with orbital phase. So if, for instance, this is the case of HAT-P-2b, the time offset between optical and near-UV transits can span through a larger interval (e.g., $0.56$ to $1.1$~h, for the conditions assumed in Fig.~\ref{fig.delta-t} and $B_*=1$~G). Similarly, the variation in $f_c$ can be more significant than the variation we estimated individually for the wind scenario or the confined scenario. 

Shock variations as caused by azimuthal variations in the coronal material can be found in both circularised and eccentric orbits. We note that WASP-12b (which is currently believed to be in a circular orbit) will also experience variations in the shock if the planet's orbit takes it through both regions of confined plasma and also wind plasma \citepalias{paper1}. In this case, azimuthal variations in the coronal material could explain the most recent results from Haswell et al. (in prep.), which show that the time-offset between the beginnings of optical and near-UV transits varied from observations done in Sep/2009 and Apr/2010.

We also note that, a similar effect could be observed in the case that the coronal material is axisymmetric (e.g., in an aligned stellar dipole), but the orbital plane and the stellar equator are not coplanar (non-null planetary obliquity).

\subsection{Case 3: Time-dependent Stellar Magnetic Field}
The last case we investigate in this paper is when the stellar magnetic field is intrinsically time-dependent, which generates variations in the stellar wind. Stellar magnetic field variations can occur on different timescales, such as short timescales due to flares or coronal mass ejections (CMEs) and large ones as due to stellar magnetic cycles. 

When a CME hits a planet, it increases the density and velocity of the surrounding material, which may cause significant compression in the size of the planet's magnetosphere. In this case, the density contrast increases, which is favourable for shock detection. As the average duration of a solar CME at a distance $6$--$10~R_\odot$ is about $8~$h \citep{Khodachenko2007}, the density increase effect caused by a CME is transient and may be easily missed, except for the case of planets orbiting young, magnetically active stars for which the magnitude and frequency of flares and coronal mass ejections may be much higher than for the present-day Sun.

Although in the case of solar-like stars we may expect a magnetic cycle on a timescale of decades (compared to $11$~yr for the Sun), the recent detection of a $2$-year cycle on the planet-hosting star $\tau$~Bootis \citep{fares2009} may indicate that shorter cycles are possible. Therefore, variations in the magnetosphere of the planet due to polarity reversals of the stellar magnetic field should occur on timescales of magnetic cycles of the host star.

\section{Summary and Conclusions}\label{sec.conclusions}
In this paper, we investigated time-dependent effects in the asymmetry of planetary transit light curves. As suggested by us, the presence of a planetary bow shock, formed from the interaction with the stellar coronal material, may be able to absorb stellar near-UV photons, causing the transit light curve in the near-UV to differ from the optical one. Our investigation of time-dependent effects in shock formation is motivated by preliminary results from a second set of near-UV observations of WASP-12b obtained in Apr/2010 (Haswell et al. in prep.). As compared to the first set from Sep/2009, it was found that the early ingress observed in Apr/2010 started before the early ingress observed in Sep/2009. A possible explanation for this may be that the size of the planet's magnetosphere changed, therefore changing the location of the bow shock around the planet.

We investigated three different scenarios that could cause the size of the planetary magnetosphere to change. In the first scenario, we considered the case of an axisymmetric stationary stellar corona and an eccentric planet whose orbital spin is parallel to the stellar axis of symmetry. Because of the planet's eccentricity, the coronal material interacting with the planet has different characteristics depending on the planetary orbital phase. Therefore, variations in the size of the planet's magnetosphere are modulated by the orbital phase. These variations can not be detected by observations conducted at one single orbital phase, which is, in general, the case for transit observations. 

However, observations that probe the size of the planet's magnetosphere and that can be conducted at any orbital phase should be modulated by the orbital phase. This is the case for observations searching for non-thermal planetary radio emission due to the interaction of the planet with the stellar material. This interaction limits the size of the planetary magnetosphere, which then constrains the size of the planet's area that is open to the stellar wind. We showed that the frequency of radio emission coming from this region will be modulated by the orbital phase, although its amplitude is at most a few MHz centred at $\sim 37$~MHz for the systems investigated in this paper. For this calculation, we assumed that the magnetic field intensity of the planet is $B_p=14$~G and of the star is $B_*=1$~G and that the stellar coronal material is at temperature $2\times 10^6~$K and has a base density of $10^8$~cm$^{-3}$.

In the second scenario investigating changes on the size of the planetary magnetosphere, we still considered that the stellar corona is steady, but allowed non-axisymmetry so that the planet shocks with material of different characteristics during its orbit. In this case, differences in the surrounding material will cause variation on the size of the planet's magnetosphere, and therefore on the stand-off distance observed during transit observations. The resultant shock variation can occur both in circularised and eccentric systems. Because the stellar rotation period in general differs from the orbital period, a series of transit observations can probe different stellar material. This is the case, for instance, if the star has an oblique magnetosphere and the planetary orbit takes the planet through regions of confined and expanding stellar material.

In the third scenario investigated, we considered time-dependent intrinsic variations of the stellar magnetism, such as due to CMEs and stellar magnetic cycles. Although the impact of a CME on a planet increases the local density surrounding the planet, its effect is transient and may not be captured in transit observations, except for the case of planets orbiting young, magnetically active stars for which the magnitude and frequency of flares and coronal mass ejections may be much higher than for the present-day Sun.

Our results show that, only in the case of a circularised planet orbiting a perfectly aligned star (axisymmetric corona where the stellar rotation axis is parallel to orbital axis), a bow shock surrounding a planet should be steady. Therefore, for systems where a shock is detectable through near-UV transit light curve observations, shock variations should be a common occurrence, seen in time-variability of near-UV transits.

\section*{Acknowledgements}
The authors acknowledge support from STFC.

\bsp
\label{lastpage}

\end{document}